\documentclass[a4paper,fleqn]{cas-dc}
\usepackage[numbers]{natbib}
\usepackage{multirow}
\usepackage[utf8]{inputenc}
\usepackage{graphicx}
\usepackage{hyperref}
\usepackage{epstopdf}
\usepackage{adjustbox}
\usepackage{float}
\usepackage{cancel}
\usepackage{soul}
\usepackage[normalem]{ulem} 
\def\tsc#1{\csdef{#1}{\textsc{\lowercase{#1}}\xspace}}
\tsc{WGM}
\tsc{QE}
\tsc{EP}
\tsc{PMS}
\tsc{BEC}
\tsc{DE}

\DeclareUnicodeCharacter{2212}{-}
\begin{document}
\let\WriteBookmarks\relax
\def\floatpagepagefraction{1}
\def\textpagefraction{.001}
\shorttitle{Shape anisotropy effect on magnetization reversal induced by linear down chirp pulse}
\shortauthors{Z. K. Juthy et~al.}

\title [mode = title]{Shape anisotropy effect on magnetization reversal induced by linear down chirp pulse}
\author[1]{Z. K. Juthy}[orcid=0000-0002-4168-8695]
\author[1]{M. A. J. Pikul}[orcid=0000-0002-7500-1572]
\author[1]{M. A. S. Akanda}[orcid=0000-0002-6742-2158]
\author[1]{M. T. Islam}[orcid=0000-0002-3846-4009]
\cormark[1]
\ead{torikul@phy.ku.ac.bd}
\address[1]{Physics Discipline, Khulna University, Khulna 9208, Bangladesh}
\cortext[cor1]{Corresponding authors: }

\begin{abstract}
We investigate the influence of shape anisotropy on the magnetization reversal of a single-domain magnetic nanoparticle driven by a circularly polarized linear down-chirp microwave field pulse (DCMP). Based on the Landau-Lifshitz-Gilbert equation, numerical results show that the three controlling parameters of DCMP, namely, microwave amplitude, initial frequency and chirp rate, decrease with the increase of shape anisotropy. For certain shape anisotropy, the reversal time significantly reduces. These findings are related to the competition of shape anisotropy and uniaxial magnetocrystalline  anisotropy and thus to the height of energy barrier which separates the two stable states. The result of damping dependence of magnetization reversal indicates that for a certain sample shape, there exists an optimal damping situation at which magnetization is fastest. Moreover, it is also shown that the required microwave field amplitude can be lowered by applying the spin-polarized current simultaneously. The usage of an optimum combination of both microwave field pulse and current is suggested to achieve cost efficiency and faster switching. So these findings may provide the knowledge to fabricate the shape of a single domain nanoparticle for the fast and power-efficient magnetic data storage device.
\end{abstract}


\maketitle

\section{Introduction}
Obtaining fast and energy-efficient magnetization reversal of single-domain of the perpendicularly magnetized nanoparticle is an interesting issue owing to its potential application in high-density data storage devices \cite{sun2000,woods2001,zitoun2002} and rapid data access \cite{hillebrands2003}.
For high density, high thermal stability and low error rate in device application, high anisotropy materials \cite{mangin2006} are needed. But it is a challenge to find out a way with low energy to achieve the fastest magnetization reversal for high-anisotropy magnetic
nanoparticle. Over the past two decades, several controlling parameters$\slash$driving forces are discovered to achieve fastest magnetization reversal with low cost. Namely, magnetization reversal using a constant magnetic field \cite{hubert1998,sun2005}, reversal by microwave field of constant frequency or time dependent frequency, either with or without a polarized electric current~\cite{bertotti2001,sunz2006,denisov2006,okamoto2008,zhu2010,thirion2003,rivkin2006,wangc2009,barros2011,barrose2013,tanaka2013,klughertzx2014} and by spin transfer torque (STT) or spin orbit torque (SOT)~\cite{slonczewski1996,berger1996,tsoi1998,Katine2000,Waintal2000,sun2000a,suns2003,stiles2002,bazaliys2004,koch2004,wetzels2006,manchon2008,miron2010,miron2011,liu2012}. All methods mentioned above are suffering from specific drawbacks \cite{hubert1998,grollier2003,morise2005,taniguchi2008,SUZUKI200993,zsun2006,wang2007,wange2008}. For example, magnetic field or microwave field driven magnetization is not energy efficient and fast as expected. In case of current (by STT) or SOT driven, the higher threshold current requirement is a bottleneck. Later on, researchers are digressed to employ the microwave chirped pulses (microwave with time-dependent frequency) which induce fast and energy-efficient magnetization reversal \cite{thirion2003,rivkin2006,wangc2009,barros2011,barrose2013,klughertzx2014} but still the required field amplitude, initial frequency and chirp rate are not small as desired practically. 
Recent studies \cite{islam2018,islam2021} have demonstrated that the fast magnetization reversal of a cubic nanoparticle (i.e., with zero demagnetization$\slash$shape anisotropy) obtained by the linear down chirp microwave pulse (DCMP). The physical mechanism of this model is that the DCMP (whose frequency linearly decreases from initial $+f_0$ to final $-f_0$) stimulates microwave energy absorption (emission) by the magnetic moment before (after) crossing the energy barrier. The efficiency of the microwave energy absorption$\slash$emission depends on how closely the microwave frequency and magnetization intrinsic frequency match during reversal. This frequency-matching is related to the energy barrier \cite{yoo2008e,dubowik1996}. However, in small-scale nanoparticle, demagnetization field$\slash$shape anisotropy (which was excluded previously) has a great importance \cite{skomski2003n,jamet2001m,yoo2008e} because, it interacts$\slash$opposes the dominated uniaxial magnetocrystalline anisotropy and thus reduces the energy barrier between two stable states. So, the magnetization reversal of a nanoparticle with finite shape anisotropy might be faster and energy efficient. In addition, fabricating perfect cubic nanoparticle (i.e., with zero demagnetization field) is a challenging issue from practical point of view. For the most device applications, the desired microwave pulse (DCMP) should be with smaller amplitude, initial frequency and low chirp rate.

Therefore, this study focuses on how the shape anisotropy ${H}_\text{shape}$ affects the energy consumption i.e., the amplitude ${H}_\text{mw}$, initial frequency $f_o$ and pulse chirp rate $\eta$ of DCMP and the magnetization switching time $t_s$. Interestingly, we found that the parameters of DCMP i.e., $H_\text{mw}$, $f_o$, and $\eta$ decrease with the increase of shape anisotropy ${H}_\text{shape}$. In addition, there is a critical ${H}_\text{shape}$ or cross-sectional area $A$, for which or larger, $t_s$ significantly reduces to 0.346 ns which is close to the theoretical limit \cite{wang2007}. For instance, in case of a sample of cross-sectional area $32 \times 32 $ nm$^2$, or $H_\text{shape} = 0.598$ T, we estimated $t_s=0.346$ ns, $H_\text{mw}=0.0187$ T, $f_o=5.2$ GHz, and optimal $\eta= 24.96$ $\text{\:ns}^{-2}$ which are significantly smaller than $t_s=0.56$ ns, $H_\text{mw}=0.045$ T, $f_o=21$ GHz, and optimal $\eta= 67.2$ $\text{\:ns}^{-2}$ for the cubic shaped or zero $H_\text{shape}$. The physical reason of these findings is related to the competition between shape anisotropy and the uniaxial magnetocrystalline anisotropy. The findings of damping dependence of magnetization reversal indicate that, for a specific $A$ or $H_\text{shape}$, there exists a optimal damping situation. Moreover, this study shows that the required microwave field amplitude of DCMP can be lowered by applying direct current simultaneously along with DCMP. Therefore, an optimum combination of microwave field pulse and current could be applied to achieve faster and power efficient magnetization reversal. So these findings might be significant to design the shape of single domain nanoparticle to produce the low-energy cost magnetic storage device with high speed data processing.

\section{Analytical Model and Method}
\begin{figure}
	\includegraphics[width=85mm]{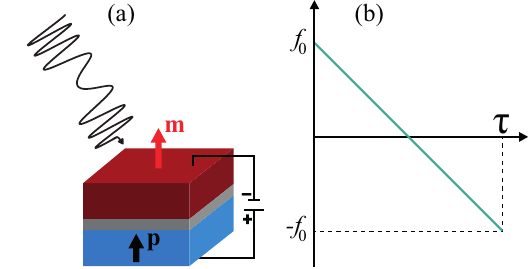}
    \caption{\label{Fig1}(a) Schematic diagram of the system in where $\mathbf{m}$ and $\mathbf{p}$ denote the unit vectors of the magnetization of free layer and fixed layer, respectively. A linear down-chirp microwave field and an electric current are applied on a single domain nanoparticle. (b) The frequency profile illustrates (sweeping from $+f_0$ to $-f_0$) the linear down-chirp microwave field.}
\end{figure}

We consider a spin valve system that consists of a free and fixed ferromagnetic layers by sandwiching a nonmagnetic spacer, as shown schematically in Figure \ref{Fig1}(a). The free layer is a square nanoparticle of area $A$ and thickness $d$. The magnetization direction of the fixed and free layer is along $\hat{\mathbf{z}}$ direction, i.e., both are perpendicularly magnetized. The magnetization of the free layer can be treated as a macrospin represented by the unit moment $\mathbf{m}$ with total magnetization $AdM_s$, where $M_{s}$ represents the saturation magnetization. The demagnetization field, which generated by the magnetization of nanoparticle, can be approximated by an easy-plane shape anisotropy. The demagnetization field$\slash$shape anisotropy field is expressed as $\mathbf{H}_\text{shape}=H_\text{shape}m_z \hat{\mathbf{z}}$ such that $H_\text{shape}=
-\mu_0(N_z-N_x)M_s$, is the shape anisotropy coefficient, where $N_z$ and $N_x$ are demagnetization factors \cite{dubowik1996} and $\mu_0 = 4\pi \times 10^{-7} \text{\:N}/\text{A}^2$ is the vacuum magnetic permeability.
 Because of the strong uniaxial magnetocrystalline anisotropy $\mathbf{H}_\text{ani}=H_\text{ani}m_z \hat{\mathbf{z}}$, the magnetization of the free layer prefers to stay in two ground states, i.e., $\mathbf{m}$ parallel to $\hat{\mathbf{z}}$ and $-\hat{\mathbf{z}}$.

Under the circularly polarized DCMP and spin polarized current, the magnetization dynamics $\mathbf{m}$ is governed by the Landau-Lifshitz-Gilbert (LLG) equation \cite{sun2005,sunz2006,taniguchi2008,gilbert2004p}
\begin{equation}
    \frac{d\mathbf{m}}{dt} = - \gamma \mathbf{m} \times \mathbf{H}_{\text{eff}} + \alpha \mathbf{m} \times \frac{d\mathbf{m}}{dt} -\gamma h_s \mathbf{m} \times (\mathbf{p} \times \mathbf{m})
\end{equation}
where $\alpha$ is the dimensionless Gilbert damping constant and, $\gamma$ is gyromagnetic ratio. $\mathbf{H}_\text{eff}$ is the total effective field which contains the microwave magnetic field $\mathbf{H}_\text{mw}$, the exchange field $\frac{2A}{M_s}\nabla^2 \mathbf{m}$ where $A$ is the exchange stiffness constant, and the effective anisotropy field $\mathbf{H}_\text{k}$ along $z$ direction, i.e., $\mathbf{H}_\text{k} = H_k m_z\hat{\mathbf{z}}$. $h_s$ is the strength of spin transfer torque(STT),
\begin{equation}
    h_s= \frac{\hbar P J}{2e \mu_o M_s d}
\end{equation}
where $J$, $e$, $\hbar$, $P$, $\mu_0$ and $d$ denote the current density, electron charge, the Planck's constant, spin polarization of current, the vacuum permeability, and thickness of the free layer, respectively.

The $\mathbf{H}_\text{k}$ has two components: an uniaxial magnetocrystalline anisotropy $\mathbf{H}_\text{ani}$ and a shape anisotropy $\mathbf{H}_\text{shape}$ field which can be expressed as $\mathbf{H}_\text{k} = \mathbf{H}_\text{ani} + \mathbf{H}_\text{shape} = [H_\text{ani} - \mu_0 (N_z-N_x)M_\text{s}] m_{z} \hat{\mathbf{z}}$. Thus, the resonant frequency of the nanoparticle can be found from the Kittel formula $f_0=\frac{\gamma} {2\pi}[H_\text{ani} - \mu_0 (N_z-N_x)M_\text{s}].$

\begin{figure*}
\centering
	\includegraphics[width=150mm]{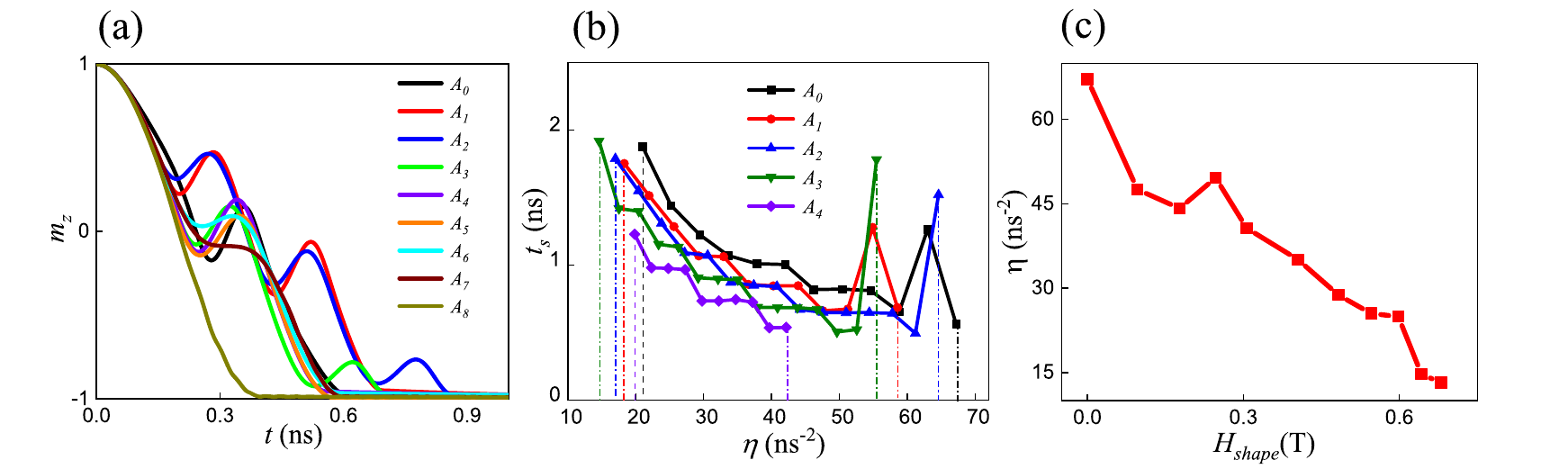}
    \caption{\label{fig:Fig2} Model parameters of the free layer, $M_\text{s}$ = $10^6$ A/m, $A_\text{ex} = 13 \times 10^{-12} \text{\:J}/\text{m}$, $H_\text{k}$ = $0.75$ T, $\gamma$ = $1.76\times 10^{11}$ rad/(T$\cdot$s), and $\alpha= 0.01$. (a) Temporal evolution of $m_z$ induced by DCMP with $H_\text{mw}=$ 0.045 T for the samples of different $A$. (b) The dependence of switching time $t_s$ on the chirp rate $\eta$ for different cross-sectional areas while keeping the microwave field amplitudes $H_\text{mw}$ constant. The vertical dashed lines indicate the lower and upper limits of $\eta$ for magnetization switching. (c) The dependence of chirp rate $\eta$ on shape anisotropy $H_\text{shape}$ for switching time within 1 ns.}
\end{figure*}

\begin{figure}
\centering
	\includegraphics[width=85mm]{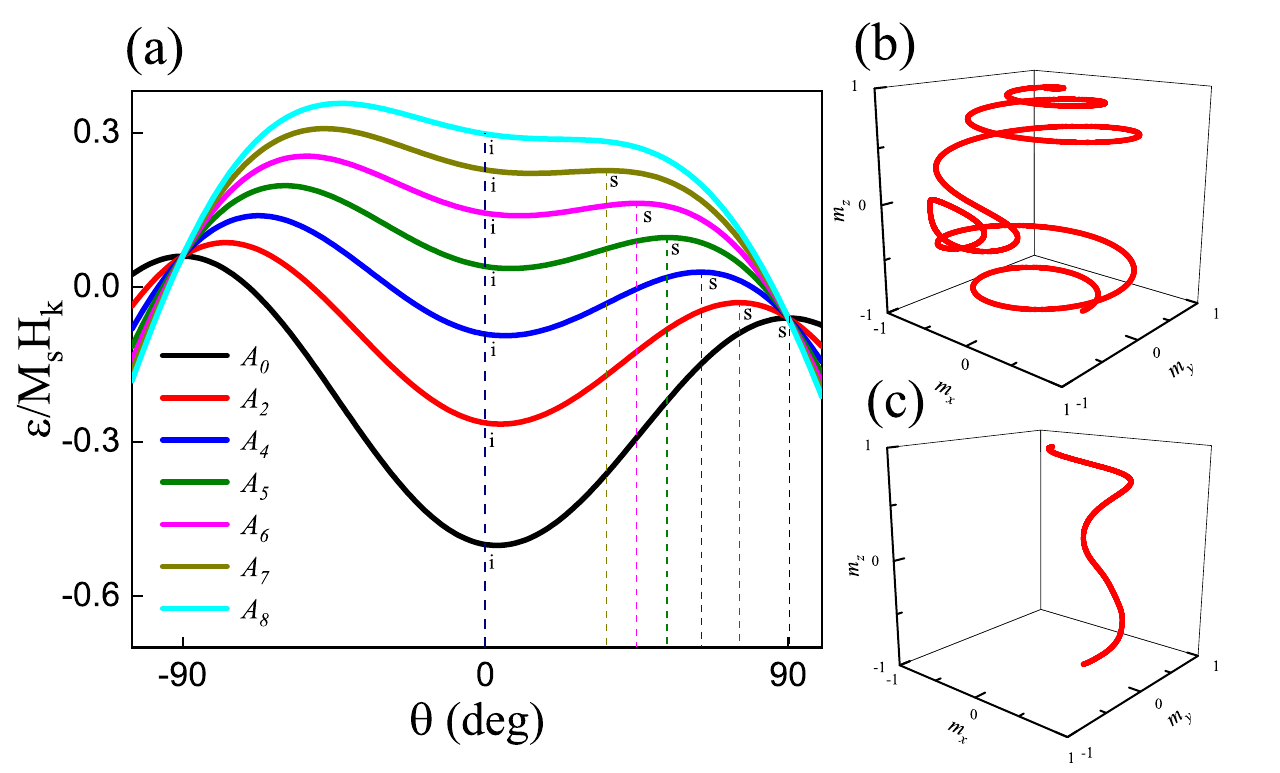}
    \caption{\label{fig:Fig3} (a) The potential $\varepsilon$ along the line $\phi$ = 0 for the samples of different $A$. Angles $\theta$ and $\phi$ are defined as $\cos^{-1}m_z$ and $ \tan^{-1}(m_x/m_y)$. The symbols i and s represent the initial state and the saddle point, respectively. Magnetization reversal trajectories of (b) $A_0=8 \times 8 $ nm$^2$ and (c) $A_8=32 \times 32 $ nm$^2$.}
\end{figure}
We first apply only the microwave field-driven magnetization reversal, i.e., without electric current (SST) term, the rate of energy change is expressed as
\begin{align}
    \Dot{\epsilon} = - \alpha \gamma \left| \mathbf{m} \times \mathbf{H}_\text{eff} \right|^2 - \mathbf{m} \cdot \Dot{\mathbf{H}}_\text{mw}
    \label{eng}
\end{align}
Since damping is positive, so, the first term is always negative. For time-dependent external microwave field, depending on the angle between the instantaneous magnetization $\mathbf{m}$ and $\dot{\mathbf{H}}_\text{mw}$, the second term can be either positive or negative, in other words, the microwave field pulse can trigger stimulated energy absorption or emission.

It has been shown that the fast magnetization reversal is achieved by DCMP \cite{islam2018} with the physical picture: the frequency of DCMP matches the frequency change of magnetization precession. Hence, it triggers stimulated microwave absorption (emission) by (from) the magnetization before (after) it crosses over the energy barrier. 

The microwave field that we have applied takes the form,
\begin{equation}
 \mathbf{H}_\text{mw} = H_\text{mw} \left[ \cos\phi(t) \hat{\mathbf{x}} + \sin\phi(t) \hat{\mathbf{y}}\right]
\end{equation}
where $ H_\text{mw}$ is the amplitude of the microwave field and $\phi(t)$ is the phase. In this paper, we use a DCMP whose instantaneous frequency is defined as $f(t) = \frac{1}{2\pi} \frac{d\phi}{dt}$, which linearly decreases with time at constant rate $\eta$ (in units of $\text{ns}^{-2}$) as shown in  Figure \ref{Fig1}(b). The phase $\phi(t)$ and the instantaneous frequency $f(t)$ are given by,
\begin{align}
\phi(t) = 2 \pi (f_0 t-\frac{\eta}{2} t^2 ) ; \ \  \ 
f(t) = f_0 -\eta t
\end{align}
where $f_o $ is the initial frequency at $t=0$. The duration of the microwave pulse is $\tau = \frac{2 f_o}{\eta}$ which means the final frequency is $-f_o$. According to the applied DCMP, the second term of Eq. \eqref{eng} can be expressed as
\begin{align}
   \dot{ \varepsilon }
    = - H_{mw} \sin\theta(t) \sin\Phi(t) \frac{d \phi}{dt}
    \label{eq:Io}
\end{align}
where $\Phi(t)$ is the angle between $\mathbf{m}_t$ (the in-plane component of $\mathbf{m}$ and $\mathbf{H}_\text{mw}$).
Therefore, the microwave field pulse can trigger the stimulated energy absorption (for $-\Phi(t)$) which is occurred before crossing the energy barrier and emission (for $\Phi(t)$) after crossing the energy barrier.

\begin{table*}[width=2\linewidth,cols=4,pos=h]
\caption{Simulated values of the controlling parameters for different samples of cross-sectional area $A$}\label{tab:table1}
\begin{tabular*}{\tblwidth}{@{} LLLL@{} }
    \toprule
    Cross$-$sectional area & Shape anisotropy field & Natural frequency & Simulated minimal frequency \\
    $A$ (nm$^2$) & \text{$h_\text{shape}$ (T)} & $f_0$ (GHz) & $f_0$ (GHz) \\
    \midrule
    $A_1$ & 0.09606 & 18.3 & 18.3 \\
    $A_2$ & 0.17718 & 16 & 17 \\
    $A_3$ & 0.2465 & 14.1 & 14. \\
    $A_4$ & 0.3064 & 12.4 & 12.4 \\
    $A_5$ & 0.4049 & 9.6 & 10.2 \\
    $A_6$ & 0.4827 & 7.49 & 9 \\
    $A_7$ & 0.5457 & 5.72 & 7.5 \\
    $A_8$ & 0.5980 & 4.26 & 5.2 \\
    \bottomrule
\end{tabular*}
\end{table*}

For this study, the parameters of Permalloy are chosen from typical experiments on microwave-driven magnetization reversal as $M_\text{s} = 10^6 \text{\:A}/ \text{m}$, $H_\text{ani} = 0.75 \text{\:T}$, $\gamma = 1.76\times 10^{11} \text{\:rad}/(\text{T}\cdot\text{s})$, $A_\text{ex} = 13 \times 10^{-12} \text{\:J}/\text{m}$ and $\alpha = 0.01$. Although, the demonstrated strategy and findings would still be valid for other materials. The cell size used throughout this study is $(2 \times 2 \times 2) \text{\:nm}^3$. We solved the LLG equation numerically using the MUMAX3 package \cite{vansteenkiste2014} for time-dependent circularly polarized microwave field. We consider the switching time window of $1 \text{\:ns}$ and least magnetization reversal $m_z = - 0.9$.

\section{Numerical Results}
In a small-scale nanoparticle, demagnetization field or shape anisotropy field $H_\text{shape}$ is an important property to take into account in the investigation of the magnetization reversal.
In order to approximate demagnetization field$\slash$shape anisotropy $H_\text{shape}$, we choose cuboid shaped sample$\slash$nanoparticle of volume $V=Ad$, where $A$ is the cross-sectional area and $d$ is the height. To increase the demagnetization field strength or $H_\text{shape}$, $A$ is gradually increased while the thickness $d=8$ nm is kept fixed. Specifically, we intend to investigate the magnetization reversal of the samples, $A_1=10 \times 10$, $A_2=12 \times 12 $, $A_3=14 \times 14$, $A_4=16 \times 16$, $A_5=20 \times 20 $, $A_6=24 \times 24$, $A_7=28 \times 28 $, $A_8=32 \times 32 $ nm$^2$. For each sample, by finding the demagnetization factors \cite{dubowik1996} $N_z$ and $N_x$, the shape anisotropy coefficient ${H}_\text{shape} = \mu_0 (N_z-N_x)M_\text{s}$ are calculated analytically. $\mathbf{H}_\text{shape}$ actually acts along $-z$ i.e., it opposes the intrinsic easy axis anisotropy field $\mathbf{H}_\text{ani}$ and hence reduces the effective anisotropy. Therefore, for each $A$, the resonance$\slash$theoretical frequency is estimated by well-known Kittel-formula $f_0=\frac{\gamma}{2\pi}\left[H_\text{ani} - \mu_0 (N_z-N_x)M_\text{s}\right]$ as shown in the Table \ref{tab:table1}. 

To investigate the DCMP-driven magnetization reversals of nanoparticles of different $A$, by keeping $H_\text{mw} = 0.045 \text{\:T}$ and $f_0$ (close to resonant frequency corresponding to $A$ or ${H}_\text{shape}$) fixed, we estimate the optimal $\eta$ for each sample or $H_\text{shape}$. Figure \ref{fig:Fig2}(b) shows the $\eta$ dependence of $t_s$ for several samples or $H_\text{shape}$. It is noted that there exists a window of $\eta$ denoted by vertical dashed line. For a small range of $\eta$ in the right edge of window, the $t_s$ is minimal. It means that there is a flexible space to choose $\eta$. In addition, the optimal $\eta(H_\text{shape})$ decreases with $H_\text{shape}$ as shown in the Figure \ref{fig:Fig2}(c). These findings are useful in device application since it was reported that generating DCMP with higher chirp rate is a challenge \cite{islam2018}. Afterward, DCMP (with $H_\text{mw} = 0.045$, $f_0$ corresponding to $H_\text{shape}$ and optimal $\eta$($H_\text{shape}$)) driven magnetization reversals are investigate and thus the time evolutions of $m_z$ for different $H_\text{shape}$ or $A$ are shown in the Figure \ref{fig:Fig2}(a). It is found that with the increase of $H_\text{shape}$, the magnetization switching time $t_s$ remains almost same until $A_7 = 28 \times 28 $ nm$^2$ but for $A_8 \geq 32 \times 32 $ nm$^2$, the magnetization reverses very smoothly i.e., $t_s$ drops to = 0.34 $\text{\;ns}$ which is close to the theoretical limit (0.39 $ \text{\;ns}$) \cite{wang2007}. So, it is mentioned that there is a critical $A$ and for larger values than it, the switching time becomes significantly smaller which is an ultimate demand in device application.

\begin{figure}
    \centering
    \includegraphics[width=80mm]{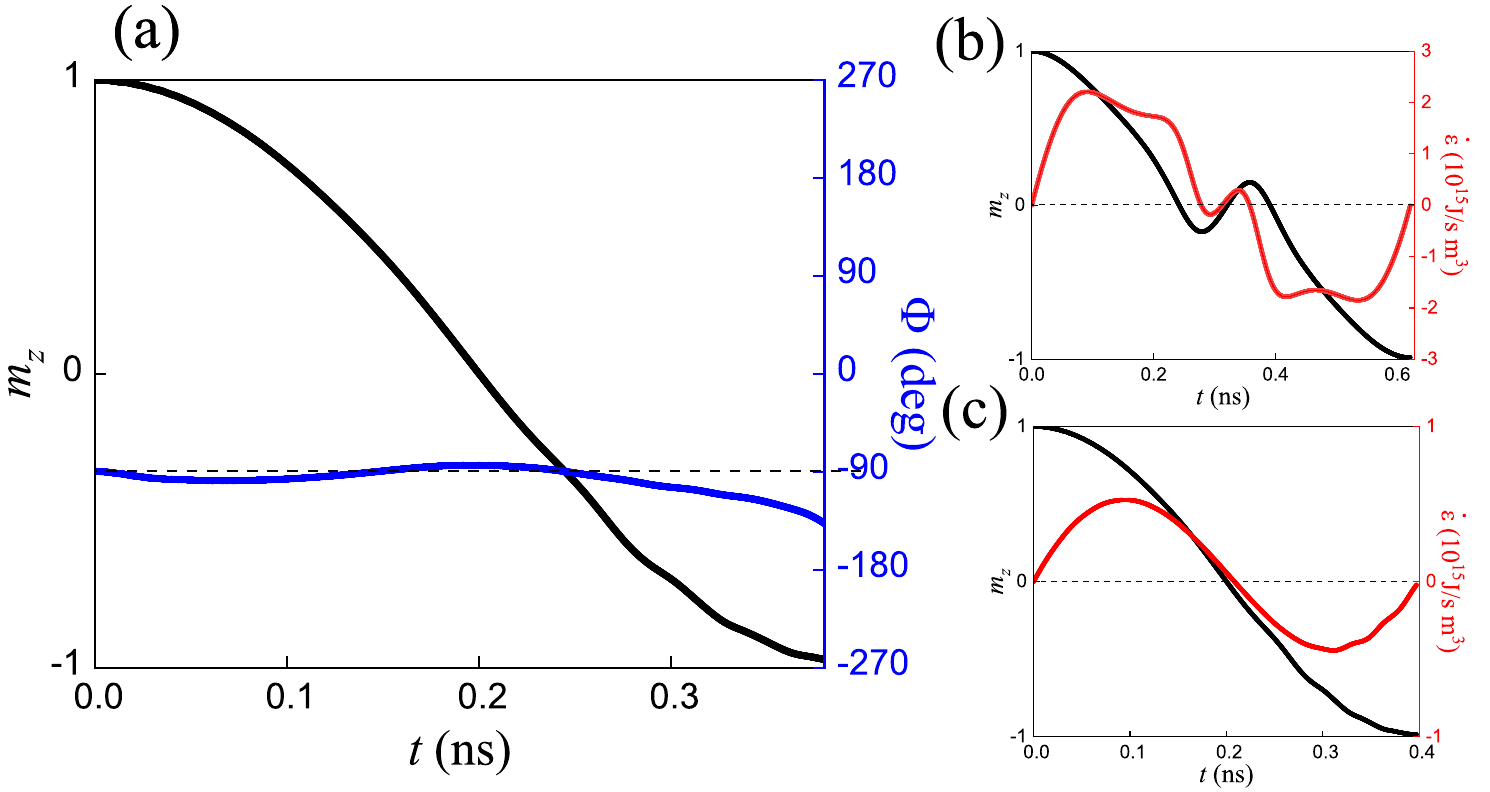}
    \caption{\label{fig:Fig5}(a) For $ A_8=32\times 32 \text{\:nm}^2$, the plot of the relative angle $\Phi$ against time (blue line) and the time dependence of $m_z$ (black line). Plots of the energy changing rate $\dot{\epsilon}$ of magnetization against time (red line) and the time dependence of $m_z$ (black line) for the cross-sectional area (b) $A_0=8\times 8 \text{\:nm}^2$ and (c) $ A_8=32\times 32 \text{\:nm}^2$ }
\end{figure}

To understand the reason why fast reversal is obtained, one can recall that the energy barrier is proportional to the effective anisotropy field \cite{zsun2006,wang2007}. Since the $\mathbf{H}_\text{shape}$ opposes the anisotropy field $\mathbf{H}_\text{ani}$ and thus reduces the height of the energy barrier (energy difference between the initial state and saddle point). This reduction of energy barrier for different $A$ is presented in the Figure \ref{fig:Fig3}(a). Therefore, the magnetization reversal becomes easier and faster because the microwave frequency and magnetization precession frequency closely matches as expected. To be more explicit, one may look at the trajectories of magnetization reversal for two cases of $A$ or $H_\text{shape}$ in Figure \ref{fig:Fig3}(b) and \ref{fig:Fig3}(c). For $A_0= 8 \times 8 $ nm$^2$ or $H_\text{shape}=0$, the magnetization takes a longer time because of some additional processions while crossing the higher energy barrier. However, for $A_8= 32 \times 32 $ nm$^2$ or $H_\text{shape}= 0.598$ T, the magnetization reversal is fairly smooth i.e., without any additional processions since the energy barrier is reduced by $H_\text{shape}$. 
For more justification, Fig \ref{fig:Fig5} (a) shows the change the angle $\Phi$ (blue) and $m_z$ (black line) with time for $A_8= 32 \times 32 $ nm$^2$. Note that the angle $\Phi$ remains almost 90$^o$ during magnetization reversal which is desired for triggering an efficient microwave energy absorption (emission) by magnetization. Thus the corresponding energy changing rate $\dot{\epsilon}$ (red), in Fig \ref{fig:Fig5}(c) is obtained with absorption and emission peaks (no distortion) which lead fast reversal. However, Fig \ref{fig:Fig5}(b), for $A_0= 8 \times 8 $ nm$^2$ i.e., for zero $H_\text{shape}$, shows the $\dot{\epsilon}$ (red line) absorption and emission peaks (distorted) indicating inefficient i.e., slower magnetization reversal (black line).

\begin{figure}
\centering
	\includegraphics[width=70mm]{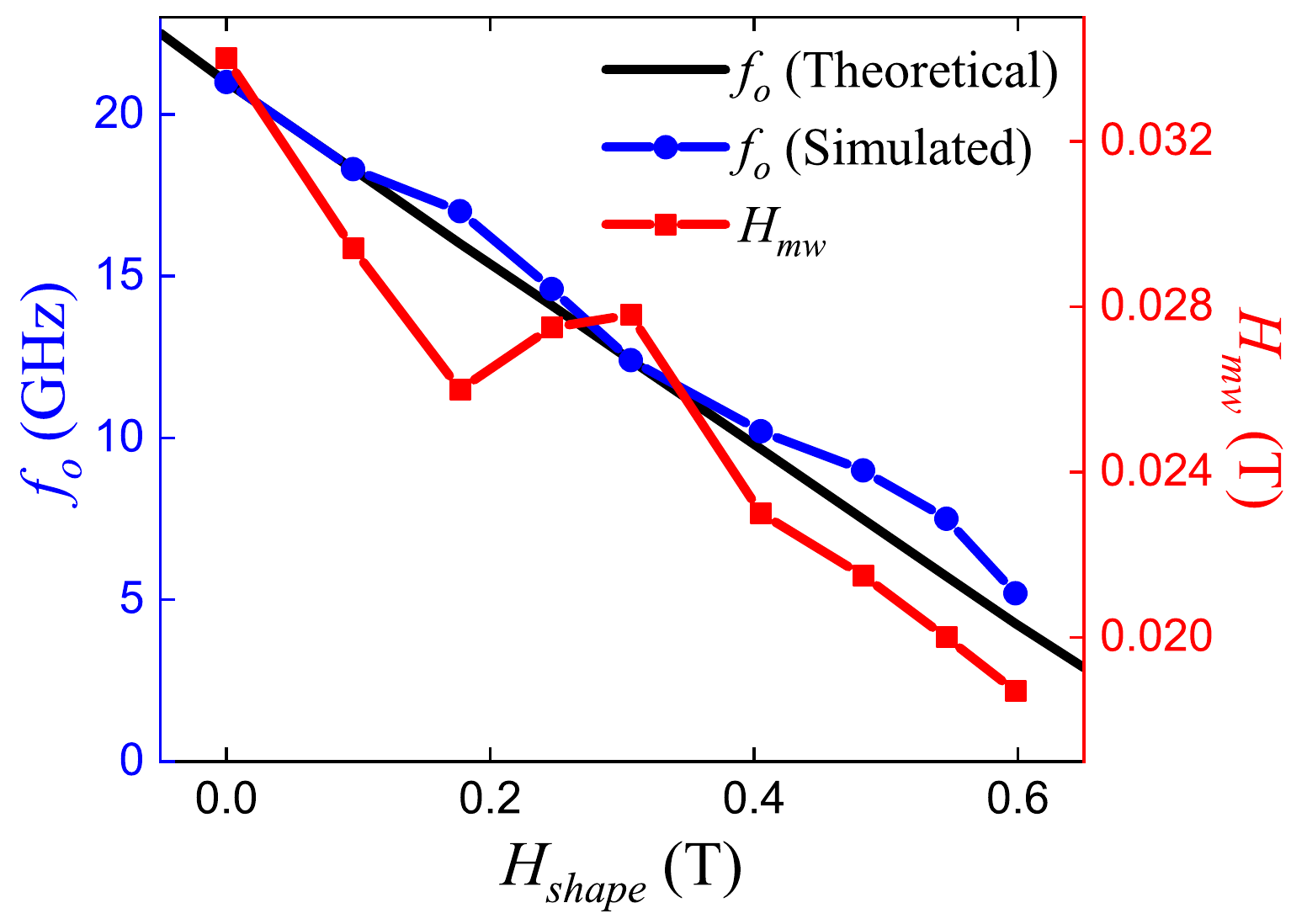}
    \caption{\label{fig:Fig4} The dependence of initial frequency $f_o$ and minimum required field amplitude $H_\text{mw}$ as a function of shape anisotropy $H_\text{shape}$ for switching time within 1 ns.}
\end{figure}

Then, for the DCMP with fixed $H_\text{mw} = 0.045$ T and optimal chirp rate $\eta(H_\text{shape})$, we obtain the fast magnetization reversal of different $A$ for a certain range of $f_0$ around each resonance$\slash$natural frequency. There exist a minimal $f_0$, i.e., minimal $f_0$ corresponding to each sample, for which switching time is lowest. Figure \ref{fig:Fig4} shows the simulated minimal frequency $f_0$ (blue circles) as a function of $H_\text{shape}$ and found that the simulated $f_0$ (blue circles) of DCMP decreases with increasing $H_\text{shape}$. These can be explained as the $H_\text{shape}$ reduces the effective anisotropy $H_\text{k}$. For further justification, the theoretically calculated frequencies (black line) are plotted in same Figure \ref{fig:Fig4} and noted that the decreasing trends are consistent as expected. Later on, by keeping the minimal $f_0(H_\text{shape})$ and optimal $\eta(H_\text{shape})$ (i.e., corresponding to different $A$ or $H_\text{shape}$) fixed, we investigate the minimally required amplitude $H_\text{mw}$(red square) as a function of $H_\text{shape}$. Figure \ref{fig:Fig4} shows that $H_\text{mw}$ also decreases with the increase of $H_\text{shape}$. This is explained by the same reason of reducing the height of the energy barrier with $H_\text{shape}$ as shown in the Figure \ref{fig:Fig2}(b). So even the smaller amplitude ${H}_\text{mw}$ can drive faster magnetization reversal.

\begin{figure}
    \centering
    \includegraphics[width=85mm]{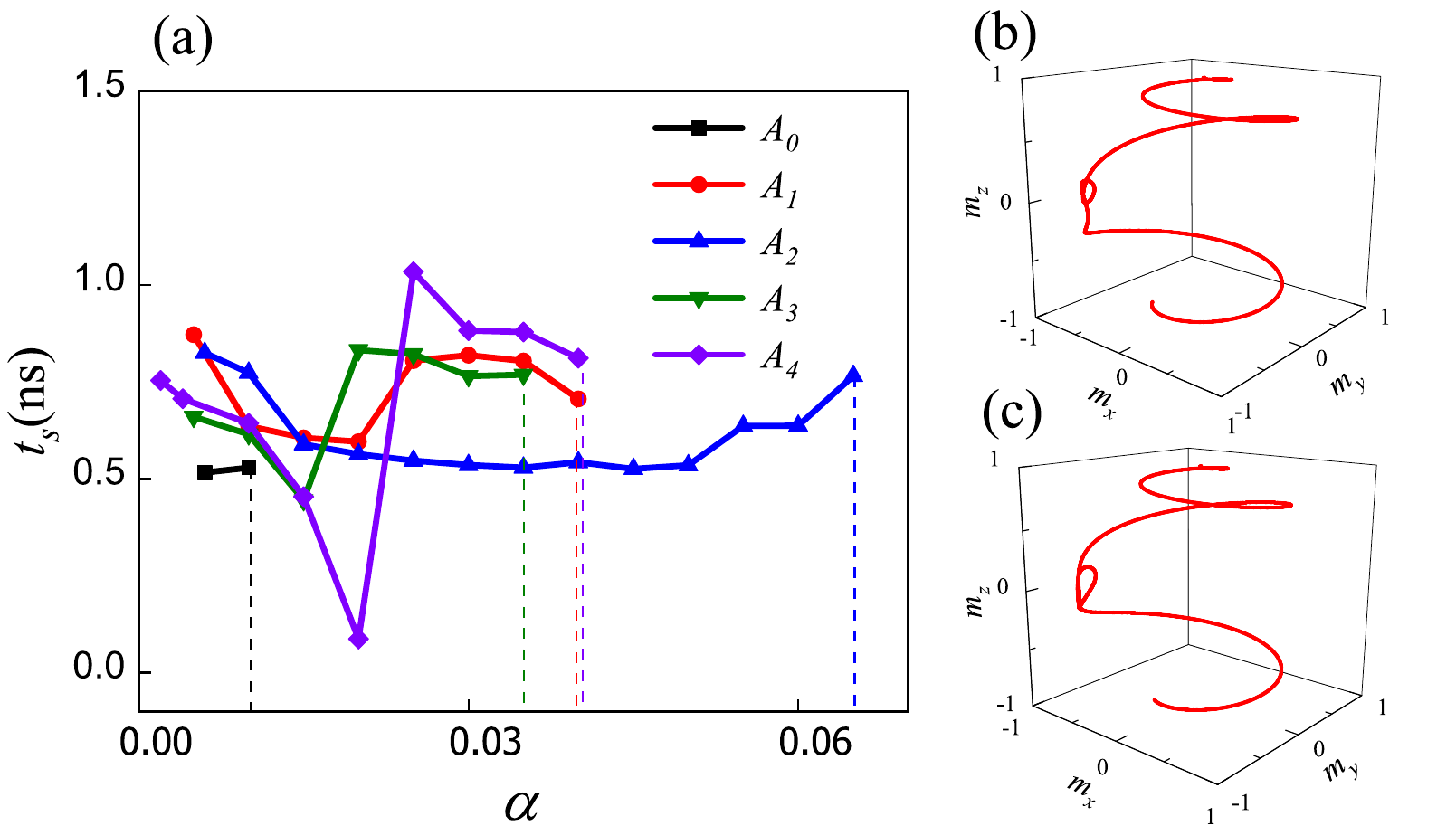}
    \caption{\label{fig:Fig6}(a) Minimal switching time for the samples of different $A$ as a function of Gilbert damping $\alpha$. Magnetization reversal trajectories of $A_3=14\times 14$ nm$^2$ for damping (b) $\alpha$ = 0.010 and (c) $\alpha = 0.017$.}
\end{figure}

The Gilbert damping coefficient $\alpha$ is an important material parameter which has a significant effect on the magnetization reversal process as well as switching time. In general, damping has two influences. First, it hinders energy absorption (cost more energy) during climbing the energy barrier for magnetization reversal \cite{wang2009c}. Second, after crossing the energy barrier, it dissipates the energy of magnetization promptly and thus it accelerates the magnetization to reach the ground$\slash$stable state \cite{uesaka2012, mao1999}. However, in the magnetization switching process, the magnetization requires to climb an energy barrier (originated from the anisotropy field) and then goes to a stable state. Ideally, for the fast and energy-efficient magnetization reversal, smaller (larger) damping ($\alpha$) before (after) cross over the energy barrier is better. Therefore, this study emphasizes finding the optimal $\alpha$ of the sample of different $A$ at which the fastest reversal is valid. Accordingly, DCMP-driven magnetization as a function of $\alpha$ for different $A$ is studied.
Figure \ref{fig:Fig6}(a) shows the Gilbert damping dependence of switching time for different $A$. It is found that for $A_0 = 8 \times 8 \text{\:nm}^2$ (uniaxial sample), the range of $\alpha$ is small but for finite shape anisotropy (biaxial sample), the range of $\alpha$ is large which is useful in device application. Moreover, for the sample of $ 14 \times 14 \text{\:nm}^2$, the switching time is lowest ($t_s=0.482$ ns) at $\alpha=0.017$. To be more explicit, Figure \ref{fig:Fig6}(b) and \ref{fig:Fig6}(c) show the trajectories of magnetization reversal for $\alpha = 0.01$ and $\alpha = 0.017$ respectively and observed that for $\alpha = 0.017$, the reversal path is shorter. This is happened because, after cross over the energy barrier, the larger $\alpha$ reduces the magnetization precession by dissipating the magnetization energy promptly and thus, it leads to fast reversal. It is indicated that there is a specific sample that has an optimal damping situation at which the magnetization reversal might be fastest.

\begin{figure}
    \centering
    \includegraphics[width=75mm]{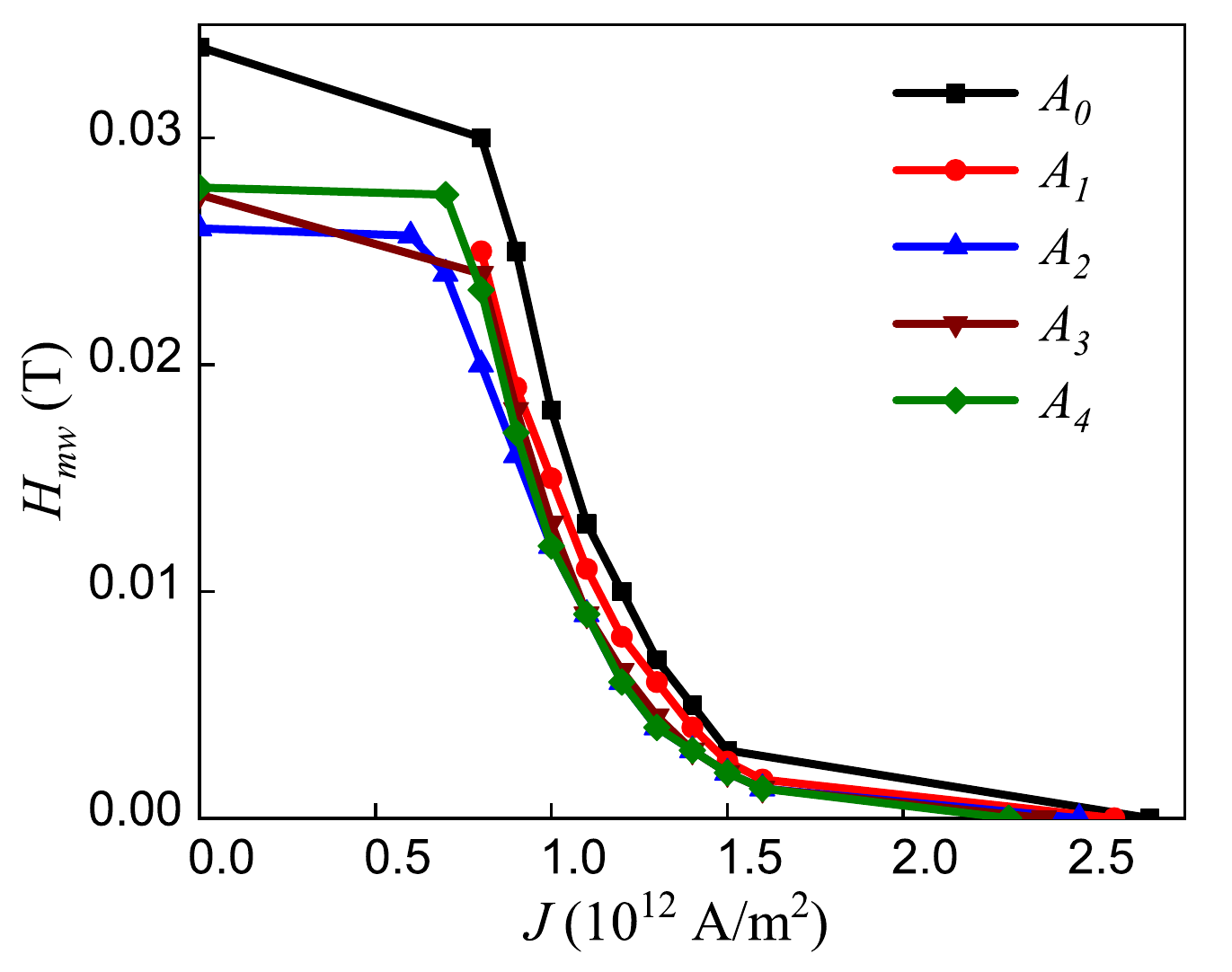}
    \caption{\label{fig:Fig7} Phase diagram, for magnetization switching within 1 ns, in terms of DCMP field amplitudes $H_\text{mw}$ and current density $J$ for the samples of different $A$.}
\end{figure}

For most device application, the desired microwave pulse (DCMP) should be of smaller amplitude $H_\text{mw}$, smaller initial frequency $f_o$ and low chirp rate $\eta$, but we have obtained, for a sample of $A = 16 \times 16 $ nm$^2$, minimal $H_\text{mw}=0.0278$ T, optimum $f_o=12.4$ GHz, and $\eta= 40.672$ $\text{\:ns}^{-2}$. In order to further reduce $H_\text{mw}$, we simultaneously apply a direct current such that it assists the magnetization reversal induced by DCMP. For DCMP contribution, we tune $H_\text{mw}$ while the minimal $f_o$ and optimal $\eta$ corresponding to different $A$ kept fixed. Figure \ref{fig:Fig7} shows the $H_\text{mw}-J$ phase diagram for the sample of different $A$ under the condition of magnetization reversal time 1 ns. As an example, for $A=16 \times 16$ nm$^2$, without DCMP, the required current to obtain fast switching gets very high i.e., $J= 2.3\times 10^{12} \text{A}/\text{m}^2 $ while without current i.e., $J = 0$, the $H_\text{mw}=0.0278$ T is large from practical point of view. Therefore, with simultaneous applying of DCMP and current, the required values of $H_\text{mw}$ and $J$ could be smaller than the above mentioned values (for individual case) which gives a large scope of practical realization of this DCMP-driven magnetization reversal strategy. 

\section{Discussion and Conclusions}
This study investigates the influence of shape anisotropy on the DCMP-driven magnetization reversal time $t_s$ and the three controlling parameters ($H_\text{mw}$, $f_o$ and $\eta$) of DCMP, and damping dependence of magnetization reversal. The shape anisotropy interestingly assists magnetization reversal since the parameters of DCMP $H_\text{mw}$, $f_o$ and $\eta$ decrease with the the increase of shape anisotropy $H_\text{shape}$. For the sample of $A_8\geq32 \times 32 $ nm$^2$, $t_s$ significantly reduces to 0.346 ns. The minimal parameters of DCMP are estimated as $H_\text{mw}=0.0187$ T, $f_o=5.2$ GHz, and optimal $\eta= 24.96$ $\text{\:ns}^{-2}$ which are useful in device application. These findings can be attributed to the shape anisotropy $H_\text{shape}$ which reduces the effective anisotropy and thus reduces the height of the energy barrier. The DCMP stimulates efficient energy absorption and emission during magnetization reversal. A recent study \cite{islam2021} reported that the temperature effect also assists the magnetization reversal, i.e., it reduces the parameters of microwave field pulse. Therefore, it is desired that, at finite temperature, the estimated values of $H_\text{mw}$, $f_o$ and $\eta$ would be decreased further (which is beyond the scope of this study). The result of damping dependence of magnetization reversal indicates that for a specific $A$ or $H_\text{shape}$, there exists an optimal damping situation. Moreover, it is also shown that the required microwave field amplitude can be lowered by applying the direct current simultaneously. The usage of an optimum combination of both microwave field pulse and current is suggested to achieve cost efficient and faster switching. So these findings may provide the knowledge to fabricate the shape of a single domain nanoparticle to generate the magnetic data storage device.

\section{Acknowledgements}
This work was supported by the Ministry of Science and Technology (Grant No. 440 EAS) and the Ministry of Education (BANBEIS, Grant No. SD2019972).

\bibliographystyle{elsarticle-num}
\bibliography{main}
\end{document}